\documentclass[conference]{IEEEtran}
\usepackage{cite}
\usepackage{amsmath,amssymb,amsfonts}
\usepackage{algorithmic}
\usepackage{graphicx}
\usepackage{textcomp}
\usepackage{hyperref}
\usepackage{subcaption}
\usepackage{xcolor}
\usepackage{booktabs}
\usepackage{multirow}

\def\BibTeX{{\rm B\kern-.05em{\sc i\kern-.025em b}\kern-.08em
    T\kern-.1667em\lower.7ex\hbox{E}\kern-.125emX}}

\definecolor{lightblue}{RGB}{32, 209, 228}
\definecolor{darkgreen}{RGB}{0, 64, 0}
\definecolor{darkyellow}{RGB}{255, 204, 0}
\definecolor{reddishpurple}{RGB}{204, 121, 167}
\definecolor{vermillion}{RGB}{213, 94, 0}
\definecolor{orange1}{RGB}{230, 159, 0}
\definecolor{blue1}{RGB}{0, 114, 178}
\definecolor{yellow1}{RGB}{240, 228, 66}
\definecolor{bluishgreen}{RGB}{0, 158, 115}

\newcommand{\tool}{V-Achilles}
\newcommand{\npma}{\texttt{npm audit}}

\begin{document}

\title{See to Believe: Using Visualization To Motivate Updating Third-party Dependencies}

\author{\IEEEauthorblockN{Chaiyong Ragkhitwetsagul\textsuperscript{*},
Vipawan Jarukitpipat\textsuperscript{*}, 
Raula Gaikovina Kula\textsuperscript{$\dagger$}, \\
Morakot Choetkiertikul\textsuperscript{*},
Klinton Chhun\textsuperscript{*},
Wachirayana Wanprasert\textsuperscript{*},
Thanwadee Sunetnanta\textsuperscript{*}}
\IEEEauthorblockA{
\textit{\textsuperscript{*}Faculty of Information and Communication Technology, Mahidol University}, Nakhon Pathom, Thailand \\
\textit{\textsuperscript{$\dagger$}Graduate School of Science and Technology, Nara Institute of Science and Technology}, Ikoma, Nara, Japan
}
}

\maketitle

\begin{abstract}
Security vulnerabilities introduced by applications using third-party dependencies are on the increase, caused by the emergence of large ecosystems of libraries such as the NPM packages for JavaScript. 
Nowadays, libraries depend on each other.
Relying on these large ecosystems thus means that vulnerable dependencies are not only direct but also indirect (transitive) dependencies.
There are automated tool supports to manage these complex dependencies but recent work still shows that developers are wary of library updates, even to fix vulnerabilities, citing that being unaware, or that the migration effort to update outweighs the decision. 

In this paper, we hypothesize that the dependency graph visualization (DGV) approach will motivate developers to update, especially when convincing developers.
To test this hypothesis, we performed a user study involving 20 participants divided equally into experimental and control groups, comparing the state-of-the-art tools with the tasks of reviewing vulnerabilities with complexities and vulnerabilities with indirect dependencies.

% This paper evaluate the effects of the existing npm dependency vulnerability visualization technique called V-Achilles on developers' decisions in updating vulnerable dependencies.
% We performed a user study involving 20 participants divided equally into experimental and control groups, comparing the state-of-the-art tools with the tasks of reviewing vulnerabilities with complexities and vulnerabilities with indirect dependencies. 
We find that 70\% of the participants who saw the visualization did re-prioritize their updates in both tasks. This is higher than the 30\% and 60\% of the participants who used the \npma~tool in both tasks, respectively. 
\end{abstract}

\section{Introduction}
\label{introduction}
The usage of third-party dependencies may lead to security vulnerabilities, which may result in unwanted access to the software.
The current software development practice is based on the ecosystem of dependencies. 
This makes the software prone to security vulnerabilities from dependencies that the software does not directly adopt~\cite{Kikas2017}\cite{Decan2018}\cite{Chinthanet2021}.
The typical response to a security vulnerability is a security advisory, which lets developers know of the security threats. 
For instance, the GitHub Advisory Database\footnote{\url{https://github.com/advisories}} contains a curated list of security vulnerabilities that have been mapped to packages tracked by GitHub. 
The security vulnerabilities are presented in the GitHub Security Advisory (GHSA) format or Common Vulnerabilities and Exposures (CVE) format along with the Common Weakness Enumeration (CWE), i.e., the type of software weaknesses that they belong to. 
The GitHub Advisory Database is sourced from the National Vulnerability Database (NVD)\footnote{\url{https://nvd.nist.gov/}}, the npm security advisories\footnote{\url{https://www.npmjs.com/advisories}}, and security advisories found and fixed by the GitHub community~\cite{GitHubSecurityAdv}.
Several automated tools are created to help the developers detecting vulnerabilities in their dependencies and to propose updates.
Two state-of-the-art tools are Dependabot \cite{Web:github_dependabot} and npm audit \cite{npmaudit}. 
% Recent work from Alfadel et al.~\cite{Alfadel:MSR2021} found that developers are highly receptive to pull requests from Dependabot.
Although having these tools, previous studies discover that the developers are still slow in updating their dependencies~\cite{Robbes2012}\cite{Hora2015}\cite{Sawant2016}\cite{Bavota2015}\cite{Ihara2017}.
The reasons are varied including not being aware of the updates, concerns about compatibility issues, and the high effort needed for adopting dependency updates~\cite{Kula2018}.
%Another work \cite{Kula2018} similarly finds that most developers were unaware of dependency updates, and also cited other issues such as the migration effort as a barrier to adopting a dependency update.

We posit that \textit{developers may re-prioritize their decisions to update}, given a visual representation of information.

% There has been tool support for developers' awareness of vulnerable dependencies and their updates.
% Two tools that have been prevalent and state-of-the-art are Dependabot \cite{Web:github_dependabot} and \npma~\cite{npmaudit}. 
%  Recent work from \cite{Alfadel:MSR2021} found that developers are highly receptive to pull requests from Dependabot.
% However, even with these tools, developers still struggle with their decision to update their dependencies. 
% % Prior studies show that developers are slow in updating their dependencies \cite{Robbes2012}\cite{Hora2015}\cite{Sawant2016}\cite{Bavota2015}\cite{Ihara2017} due to various factors such as compatibility issues, being unaware, and the migration effort outweighing the benefits.
% % Another work \cite{Kula2018} similarly finds that most developers were unaware of dependency updates, and also cited other issues such as the migration effort as a barrier to adopting a dependency update.
% % We posit that \textit{developers may re-prioritize their decisions to update}, given a visual representation of information.
% Concretely, we not only consider the direct dependencies, which are libraries that the application calls, but also we consider \textit{indirect dependencies}, which are transitive dependencies on which the libraries are dependent. 
% %Indirect dependencies can be created in different levels depending on how many dependency calls occur. 
% A thread of the direct and indirect dependencies that depend on each other can be called a \textit{chain of dependencies}. 

Jarukitpipat et al.~developed \tool~\cite{Jarukitpipat2022},  a prototype that shows a visualization (i.e., using dependency graphs) affected by vulnerability attacks.
The tool considers both the direct dependency, which is a library that the application calls, and the \textit{transitive dependencies}, which is a library that is called by a direct dependency or another transitive dependency.
%Indirect dependencies can be created in different levels depending on how many dependency calls occur. 
A previous study by Zimmerman et al.~\cite{Zimmermann2019} reports that one installed direct dependency can depend on approximately 80 transitive dependencies and the ratio seems to be increasing over time.
%Prior work~\cite{Zimmermann2019} shows that there are approximately 80 transitive dependencies per direct dependency installed, and this ratio is increasing over time. 
It is difficult for developers to detect flaws or vulnerabilities in transitive dependencies because they are invisible from the developer's point of view~\cite{Cox2019,Snyk2020}.

Nonetheless, the usefulness of the dependency vulnerability visualization to the actual software developers compared to existing techniques such as Dependabot or \npma~is still under-explored.
% This paper is an extension of the \tool~~\cite{Jarukitpipat2022} which was published as a tool demonstration paper in the Proceedings of the 37th IEEE/ACM International Conference on Automated Software Engineering (ASE '22). 
We fill the gap in this work by performing a comparison study of the dependency security analysis tools through a User Study. In detail, we compare \tool, Dependabot and \npma, using two tasks:

\textbf{Task 1: Navigating dependencies with complex graphs.} 
\textit{Methodology:} In this task, the user is given an npm project with vulnerable direct dependencies that contain several transitive dependencies. The complexity of the dependency chains, i.e., having multiple transitive dependencies, may affect the decision of the developers on updating such vulnerable direct dependencies.
    
\textit{Results:}~We found that seven out of ten developers who initially used Dependabot and later used \tool~changed their prioritization of updating vulnerable dependencies, which is more than \npma~(three out of ten). Some of the developers mentioned that they took the complexity of the dependency graph into account after using \tool.
    
\textbf{Task 2: Navigating transitive dependencies with vulnerabilities.} \\ \textit{Methodology:} In this task, the user is given an npm project with vulnerabilities in the transitive dependencies. This type of vulnerability in transitive dependencies needs to be fixed by updating the direct dependency that depends on it. This may affect the decision of the developers on updating such direct dependency. 
    
\textit{Result:}~The result in this task also shows that seven out of ten developers who initially used Dependabot and later used \tool~changed their dependency update priority, which is more than \npma~(six out of ten). After using \tool, some developers included transitive dependencies in their update decisions.

The rest of the paper is organized as follows. Section II explains the existing tool support. Section III describes our empirical study. Section IV presents the results of the empirical study. Section V discusses the related work.
Sections VI and VII discuss the threats to validity and the conclusion.

% \chaiyong{Duplicated with the previous paragraph and can be removed}
% Our results indicate using our tool, navigating through complex dependencies helps developers to prioritize their updates. For task 1, seven participants instead opted to re-prioritize based on the inter-dependencies compared to the other tools. 
% For task 2, we found that our visualization allowed users to consider direct and transitive dependency as a factor when prioritizing when to update. 
% We showed that participants who were exposed to our visualization considered the transitivity (indirect) as a factor compared to the other tools.
% Our results show that the visualization of such transitive vulnerabilities did allow participants to re-prioritize their update decisions. 

% The rest of the paper is organized as follows. Section II explains the existing tool support including GitHub Dependabot and \npma. Section III describes our \tool~prototype, its visualization concept, and its potential use cases. Section IV presents the empirical evaluation of \tool~by performing detection of vulnerable dependencies in real-world npm open source projects and evaluation of the visualization through a user study and their results. Section V discusses the results. 
% Sections VI and VII discuss the implications of \tool and the related work. Section VIII lists the threats to validity. Finally, section IX concludes the paper.

\section{Existing Tool Support}\label{sec:ExistingTools}
Currently, the two widely-used automated tools that assist developers in analyzing vulnerabilities in software projects are GitHub Dependabot and \npma.

\begin{figure}[tb]
	\centering
		\includegraphics[width=\columnwidth]{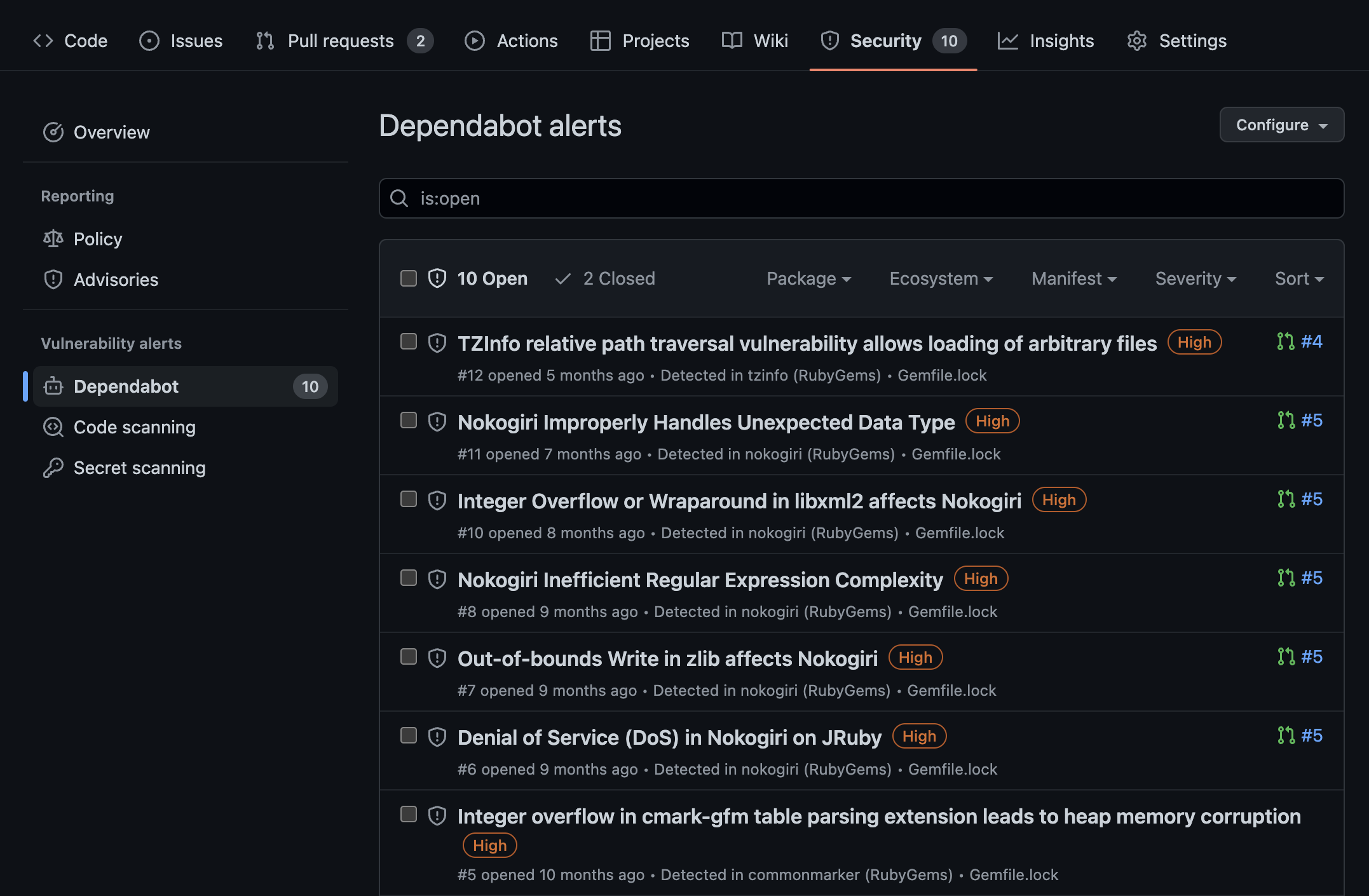}
		\caption{Dependabot Alerts}
	\label{fig:dependabot}
\end{figure}

\subsection{Dependabot}
\textit{Dependabot}\footnote{\url{https://github.com/dependabot}} is a bot in the GitHub repository that frequently analyzes the dependencies in the repository and creates alerts containing reports about the vulnerabilities in the dependencies. The tool currently supports 15 programming languages. Dependabot locates the outdated or vulnerable dependencies and then creates pull requests proposing updates for such outdated or vulnerable dependencies. The developers in the project can check the pull requests and make decisions about whether to accept the proposed dependency updates or not.
However, it has a limitation of detecting only vulnerabilities in direct dependencies\footnote{\url{https://github.com/Dependabot/Dependabot-core/issues/2640}}~\cite{dependabot2020}. Figure~\ref{fig:dependabot} shows the dependabot alerts in a GitHub project.

\begin{figure}[tb]
	\centering
	\includegraphics[width=\columnwidth]{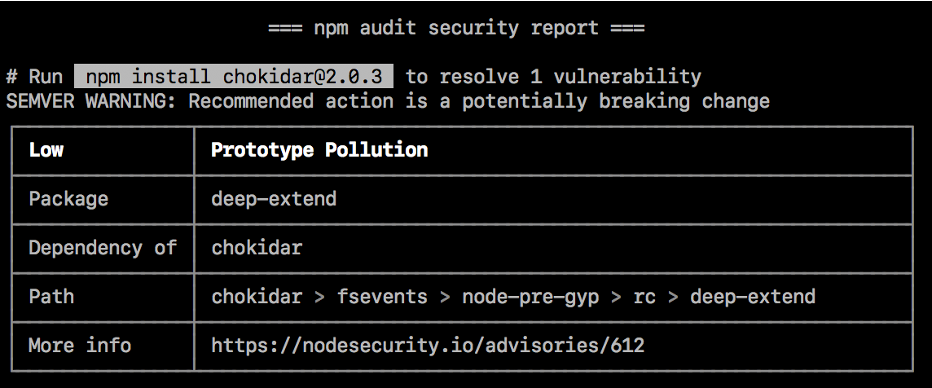}
	\caption{\npma~report}
	\label{fig:npm_audit}
\end{figure}

\subsection{npm audit}
\npma~is a part of npm, a popular package manager for the JavaScript language\footnote{\url{https://docs.npmjs.com/cli/v8/commands/npm-audit}}. It was introduced in 2018. The tool works as a command line tool. It extracts the list of JavaScript packages listed in \texttt{package.json}, i.e., the package metadata file, residing in a given npm project to the npm registry, and checks for vulnerabilities. The execution of \npma~can be performed manually by the developer or automatically when the installation of the new dependencies from npm occurs. Different from Dependabot, \npma~can report both direct and transitive dependencies in npm packages. However, it gives the report in a textual tabular format in the command line. Thus, it may be difficult to comprehend when there are many vulnerabilities in the project.
Figure~\ref{fig:npm_audit} shows the result from running \npma.

% 	The two tools can support the developers by informing them of vulnerable dependencies. Nonetheless, they are both text-based and the developers need to figure out by themselves how the vulnerable dependencies connect to other dependencies. %Also, the developers need to investigate the vulnerabilities further by themselves according to the information of CVE given in the Dependabot or npm audit report.

\subsection{V-Achilles with Dependency Graph Visualization}
\label{sec:V-Achilles}
The third tool is \tool~\cite{Jarukitpipat2022}, which adopts the visualization concept called Dependency Graph Visualization (DGV). Similar to \npma, \tool~works with only npm projects and reports vulnerabilities in both direct and transitive dependencies. We explain its visualization concept in detail below.

% \subsubsection{DGV Overview}

% The node-link dependency graph visualization that can visualize the dependencies in a given npm project. We call the approach ``DGV'' (\textbf{D}ependency \textbf{G}raph \textbf{V}isualization) and how we realize this concept as a prototype tool called \tool.

\begin{figure*}
	\centering
 \includegraphics[width=0.7\textwidth]{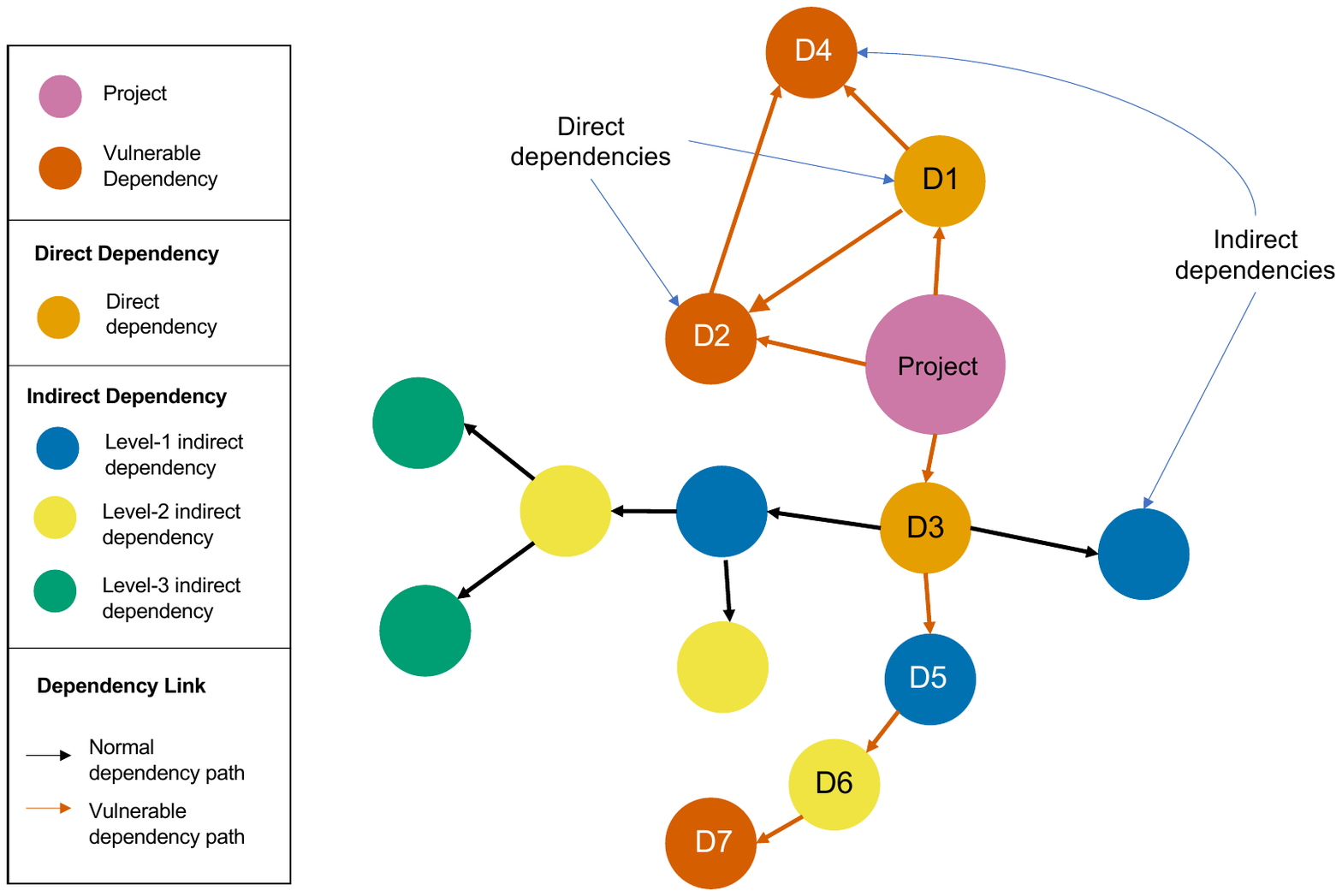}
	\caption{Dependency graph visualization with vulnerability information of \tool~\cite{Jarukitpipat2022}}
	\label{fig:viz-concept-2}
\end{figure*}

\tool~employs a color-blind safe palette in representing the dependencies~\cite{Wong2011}. The visualization contains the project node with 
\textcolor{purple}{reddish purple} color. The project node is rendered slightly larger than other nodes. The nodes that represent vulnerable dependencies are displayed in \textcolor{vermillion}{vermillion}. Normal nodes are displayed differently according to their types as follows. Direct dependencies are displayed in \textcolor{orange1}{orange}, while indirect dependencies are displayed in \textcolor{blue1}{blue}, \textcolor{yellow1}{yellow}, and \textcolor{bluishgreen}{green}. The three colors of \textcolor{blue1}{blue}, \textcolor{yellow1}{yellow}, and \textcolor{bluishgreen}{green} represent the number of hops between the node, i.e., the dependency, and the project node. The connections between the nodes, i.e., edges, are rendered in black, except the edges that connect to vulnerable dependency nodes will be displayed in \textcolor{vermillion}{vermillion}.

Figure~\ref{fig:viz-concept-2} shows how the graph visualization is applied to npm dependencies. 
In this example, we have an npm project \texttt{P} which uses 3 dependencies, \texttt{D1}, \texttt{D2}, and \texttt{D3}. Their relationship can be captured in a graph as shown in the figure. Node \texttt{P} has an edge pointing to the node \texttt{D1}, \texttt{D2}, and \texttt{D3} showing that it depends on these three dependencies. Moreover, the dependency \texttt{D1} also depends on the dependency \texttt{D2}, and another dependency \texttt{D4}. At the same time, \texttt{D2} also depends on \texttt{D4}. Thus, there are edges from the node \texttt{D1} to \texttt{D2} and to \texttt{D4} and another edge from the node \texttt{D2} to \texttt{D4}. 
We can easily identify the direct and indirect dependencies. The nodes that have the path length of 1 from the project node \texttt{P} in the graph are direct dependencies (\texttt{D1}, \texttt{D2}, \texttt{D3}). The rest are indirect dependencies (\texttt{D4}). Furthermore, we can also see the fan-out of each node, i.e., the number of dependencies that the node depends on. For example, in this figure, we can see that the project \texttt{P} has a fan-out value of 3 by having 3 outgoing edges to \texttt{D1}, \texttt{D2}, and \texttt{D3} respectively.

Although Dependabot and \npma~can report dependencies and their vulnerabilities, they still have some limitations in their visualization of the results to the developers.  Thus, the developers may not see the relationships between the dependencies and the severity of the vulnerabilities when making their decisions to update the vulnerable packages. 

In this paper, we perform a user study on the effects of DGV on developers' decisions on updating third-party dependencies compared to Dependabot and \npma.

\section{Empirical Study}
We performed an empirical study using the user study method to see the effects of the DGV provided by \tool~on the developers when they are making the decisions to update vulnerable dependencies.
The study aims to answer the following research question.

% \textbf{RQ1: To what extent are direct and indirect dependencies with vulnerabilities found in popular open-source npm projects?} Our motivation is to check how often vulnerable direct and indirect dependencies are found in open-source projects.

\textit{RQ: To what extent does our visualization influence the developer’s decision to update?} 
Our motivation is to evaluate whether or not providing relevant information in the form of visualization will cause developers to re-prioritize their update decisions compared to Dependabot and \npma.

\subsection{User Tasks in the Study}
We designed two tasks where the participants will use \tool, Dependabot, and \npma~to guide their decisions. 

\begin{figure}
		\centering
        \includegraphics[width=\columnwidth]{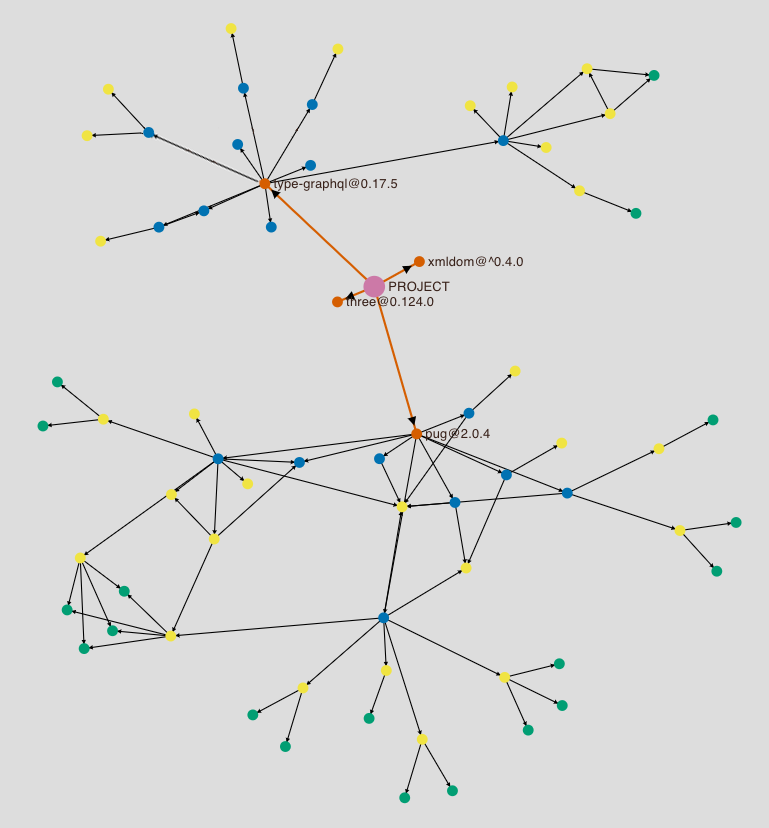}
		\caption{Dependency graph visualization of Task 1}
		\label{fig:graphtest1}
\end{figure}

\begin{table}[tb]
\centering
%\footnotesize
\caption{Dependencies in Task 1 -- Navigating dependencies with complex graphs}
\begin{tabular}{llcccc}
	\toprule
	No & Dependency & Version & Severity & Type \\
	\midrule
	1 & three & 0.124.0 & High & Simple \\
	2 & pug & 2.0.4 & High & Complex \\
	3 & xmldom & \^{}0.4.0 & Low & Simple \\
	4 & type-graphql & 0.17.5 & Low & Complex \\
	\bottomrule
\end{tabular}
\label{table:cha-teat1}
\end{table}

\paragraph{Task 1: Navigating Dependencies with Complex Graphs} 
For Task 1, we aim to test if \tool~with DGV's users will change their dependency updates when dealing with complex dependencies.
We define a project with complex dependencies as a project with multiple transitive dependencies, hence, it is depicted with multiple nodes.
This represents a situation where an update of a vulnerable dependency may cause several changes due to the changes in the transitive dependencies it depends on.
To test for this situation, we artificially created an npm project that contained four real-world vulnerable dependencies that match our definition of complex dependencies: \texttt{three}, \texttt{pug}, \texttt{xmldom}, and \texttt{type-graphql}. 

Figure \ref{fig:graphtest1} shows the visualization of the dependencies, we can see that the four selected dependencies are direct dependencies of the project. \texttt{pug} and \texttt{type-graphql} are complex dependencies. \texttt{pug} directly depends on eight other dependencies and \texttt{type-graphql} directly depends on nine other dependencies (highlighted with blue nodes). Their dependencies also depend on several other dependencies (yellow and green nodes). \texttt{three} and \texttt{xmldom} are not complex as they do not have any transitive dependencies.
Table~\ref{table:cha-teat1} shows the information on the dependencies for this task. 
We were interested in the dependencies which had different severity levels and also had different complexities. Thus, we selected two dependencies, \texttt{three} version 0.124.0 and \texttt{pug} version 2.0.4, that had high severity vulnerabilities and two dependencies, \texttt{xmldom} version range of \^{}0.4.0 and \texttt{type-graphql} version 0.17.5 that had low severity vulnerabilities. 
%Two of the dependencies, \texttt{three} and \texttt{xmldom}, did not have any dependencies while the other two dependencies, \texttt{type-graphql} and \texttt{pug}, depended on 8 transitive dependencies, i.e., they introduced complexity into the chain of dependencies. 
We cross-checked to make sure that the four selected vulnerable dependencies were reported by Dependabot, \npma, and \tool. The dependencies in this task can be categorized into 4 categories: high-severity simple, high-severity complex, low-severity simple, and low-severity complex. These four distinct categories help us to understand the reasoning behind the developers' rankings better. Since Dependabot and \npma~cannot show the dependency graph complexity, the developers using \tool~may take the complexity along with the severity into account when ranking the dependencies to update.

% To answer RQ1, we compare the differences between groups that change their priorities after being exposed to the different tools. 
% We call them the \textit{re-prioritize group} and the \textit{unchanged priority group}. 
% The re-prioritize group contains the participants whose dependency prioritization after seeing our visualization or \npma~report differs from the  initial prioritization based on the Dependabot report, i.e., they changed their ranking of the vulnerable dependencies to be fixed. 
% The unchanged priority group contains the participants whose dependency prioritization was the same between \tool~and Dependabot or \npma~and Dependabot.

\begin{figure}
    \centering
    \includegraphics[width=0.7\columnwidth]{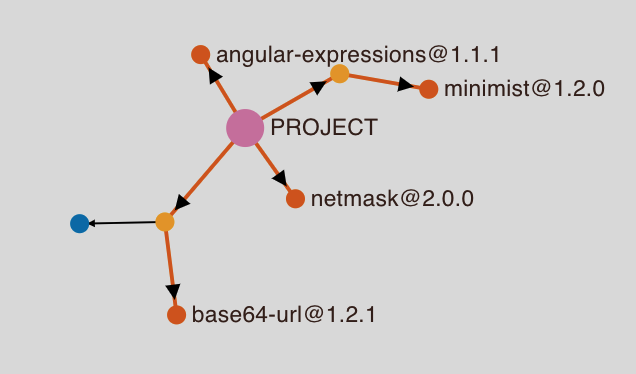}
    \caption{Dependency graph visualization of Task 2}
    \label{fig:graphtest2}
\end{figure}

\begin{table}
\centering
\caption{Dependencies in Task 2 -- Navigating transitive dependencies with vulnerabilities}
%\footnotesize
%\resizebox{\columnwidth}{!}{%
\begin{tabular}{llcccc}
	\toprule
	No & Dependency & Version & Severity & Type \\
	\midrule
	1 & netmask & 2.0.0 & High & Direct \\ 
	2 & base64-url & 1.2.1 & High & Transitive \\
	3 & angular-expressions & 1.1.1 & Low & Direct \\ 
	4 & minimist & 1.2.0 & Low & Transitive \\
	\bottomrule
\end{tabular}
%}
\label{table:cha-test2}
\end{table}

\paragraph{Task 2: Navigating Transitive Dependencies with Vulnerabilities}
For Task 2, we aim to test whether \tool's DGV which shows the types of vulnerabilities (direct or transitive) would affect the participants’ decision on dependency prioritization. This represents a situation where the security vulnerabilities are not from the direct dependencies of the project, but transitive ones.
We created another artificial npm project for this task and also made sure they could be detected by the three tools. 

The selected four dependencies include \texttt{netmask} version 2.0.0, \texttt{base64-url} version 1.2.1, \texttt{angular-expressions} version 1.1.1, and \texttt{minimist} version 1.2.0. Two dependencies, \texttt{netmask} and \texttt{base64-url} had high severity vulnerabilities, while \texttt{angular-expressions} and \texttt{minimist} had low severity vulnerabilities. 
Figure \ref{fig:graphtest2} and Table \ref{table:cha-test2} show the information on the dependencies in Task 2 and how they depend on each other. We can see from the visualization that \texttt{angular-expressions} and \texttt{netmask} are the direct dependencies of the project, while \texttt{minimist} and \texttt{base64-url} are transitive dependencies of the project.
Similarly to Task 1, they can be categorized into four groups: high-severity direct dependency, high-severity transitive dependency, low-severity direct dependency, and low-severity transitive dependency. 
Since Dependabot cannot show transitive dependencies and \npma~shows transitive dependencies using the textual format, the developers using \tool~may use the visualization of direct vs.~transitive dependencies along with their severity into account when ranking the dependencies to update.

\paragraph{Experiment Settings}

The user study follows the guidelines from Ko et al.~\cite{Ko2013}.
% 	We highlight the key steps from the guidelines, which consists of Recruitment, Selection, Consent, Procedure, Demographic measurements, Group assignment, Training, Tasks, Outcome measurements, and Debrief and compensate. 
We recruited twenty participants for the study (as shown in Table~\ref{table:participants}).
Following the between-subjects experimental design~\cite{Charness2012}, we randomly assign the participants into two groups: the control group and the experimental group. 
The control group has ten participants. The participants in this group used \npma~to analyze security vulnerabilities. For the experimental group, the ten participants used \tool~to perform the same tasks. 
We used the control group as a baseline to compare our findings with those exposed to \tool.
Table~\ref{table:participants} describes the demographics of the participants of the study.
We set the inclusion criteria for selecting the participants of this experiment as follows.
\begin{enumerate}
	\item The participants must be computer science students or full-time employees at a software development company
	\item The participants must have experience in programming for at least six months.
	\item The participants have a fundamental knowledge of software vulnerabilities and related tools.
\end{enumerate}

\begin{table}[bt]
	\caption{Participants' Demographic and Tools Assignment}
	\centering
	\footnotesize
	%\resizebox{\columnwidth}{!}{%
		\begin{tabular}{p{2cm}cp{1.5cm}p{2cm}}
			\toprule
			Group & Participants & Know Trans. Dep. & Tools Assignment \\ 
			\midrule
			 V-Achilles & E1 & No & Dependabot \\ 
                (Experimental  & E2 & Yes & followed \\ 
			Group) & E3 & Yes & by \tool\\ 
			& E4 & No \\ 
			& E5 & Yes \\
			& E6 & No \\
			& E7 & No \\ 
			& E8 & No \\
			& E9 & No \\
			& E10 & No \\
			\midrule
			 npm-audit & C1 & Yes & Dependabot \\ 
			(Control Group) & C2 & No & followed \\ 
			& C3 & Yes & by \npma \\
			& C4 & No \\
			& C5 & No \\
			& C6 & Yes \\
			& C7 & No \\
			& C8 & No \\
			& C9 & No \\
			& C10 & No \\
			\bottomrule
		\end{tabular}
	%}
	\label{table:participants}
\end{table}

Then, we performed several methods to recruit the participants. For developers, we sent an email to invite them. For students, we directly contacted the students at the Faculty of Information and Communication Technology, Mahidol University, Thailand.  Table~\ref{table:participants} shows their demographics, 
%their experience with using npm (high means more than 3 years), 
and whether they know and understand the concept of transitive dependencies along with their group assignments. Three participants were full-time developers. Six participants were master's students and eleven participants were undergraduate students. 
%Six participants were proficient in using npm and six participants knew about transitive dependencies before the study. 
The number of participants who knew transitive dependencies was the same in the two groups (three participants).

\textbf{Pre-Task Procedure}.
After getting the participant's consent, we gathered their demographic background, provided the guidelines and videos of the demonstration of the \tool, Dependabot, and \npma~tools, and guided them to perform two example tasks to train them to get used to the tools and security vulnerabilities in npm projects.
Then, we introduced the two tasks (i.e., Task 1 and Task 2) that we used for the study to the participants (the details of each task are as explained above). 
%For each task, the participants had to do the following.

\textbf{Ranking}.
As shown in Table~\ref{table:participants}, the participants in both the control and the experimental groups were given an npm project and the vulnerability report of Dependabot. 
Then, they were asked to rank the vulnerable dependencies that they thought they would fix from the first to the last. 
After that, a different tool was introduced to each group. 
For the control group, the \npma~and its report was introduced. For the experimental group, \tool~with dependency graph visualization was introduced. 
After using the given tool to explore the security vulnerabilities, the participants were asked to rank the vulnerable dependencies to fix again. 
%Finally, we exposed the control group to \tool~after they had finished using the \npma. 
%Then, they performed the dependency prioritization for the last time.

\textbf{Post-Task Interview}.
We interviewed participants after they finished using each tool for the tasks. 
After they finished prioritizing the update, we asked them 

\begin{quote}
\textit{``What criteria did you use to prioritize these vulnerability updates?''}    
\end{quote}

Notes were taken while the participants verbally explained the criteria they considered when prioritizing the vulnerability updates.

\paragraph{Analysis of the Ranking}
We compared the prioritization results that the participants used before (i.e., based on the security vulnerability report from Dependabot) and after (i.e., based on the security vulnerability report from \npma~or \tool) of the two tasks. 
This is classified as being the same or different.
Same prioritization means the ranking of the dependencies to be updated is exactly the same between using Dependabot and \tool~(or \npma). 
On the contrary, different prioritization means the ranking of the dependencies to be updated is changed.
%We also recorded whether or not the priorities are the same or different when the control group was finally gets exposed to our visualization.
%For Task 1 and Task 2 (Table~\ref{table:result-Ach-t1e} and Table \ref{table:ach-indirect-e}),
%the re-prioritize group has seven participants that changed their initial prioritization order from Dependabot after seeing our visualization. 

% Dependabot, \npma~and \tool~provide similar basic security vulnerability information to the users with some slight differences. The common information provided by the three tools includes the list of dependencies (i.e., package names) and versions, their severity, and the patched versions. \npma~and \tool~provide the chain of dependencies while Dependabot and \tool~provide CWE, CVE, and GHSA ID to the users.

% We are interested in the prioritization of the npm dependencies because the developers need to take time to adjust to new changes introduced by performing the update. Thus, by making the right prioritization (which depends on many factors such as existing packages used in the project, expertise of the developers, and understanding of the project), the developers can save their time by selecting only the update that is most important or more feasible to do within their allocated time.

\section{Results}

Our results from the user study are presented in two ways. First, we quantitatively measure the feedback from the experiment, and then we qualitatively analyze the feedback of the participants.

% \subsection{Task One: Vulnerabilities with complexities}

\begin{figure}[ht]
\begin{subfigure}{\columnwidth}
  \centering
  % include the first image
  \includegraphics[width=\columnwidth]{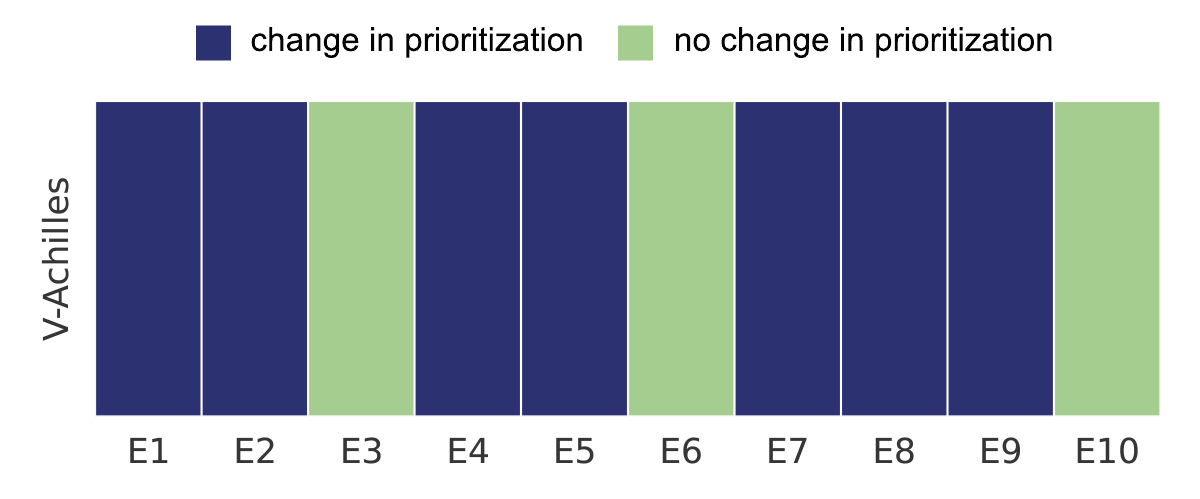}  
  \caption{Experimental Group: V-Achilles}
  \label{fig:task1_exp}
\end{subfigure}
\begin{subfigure}{\columnwidth}
  \centering
  % include the second image
  \includegraphics[width=\columnwidth]{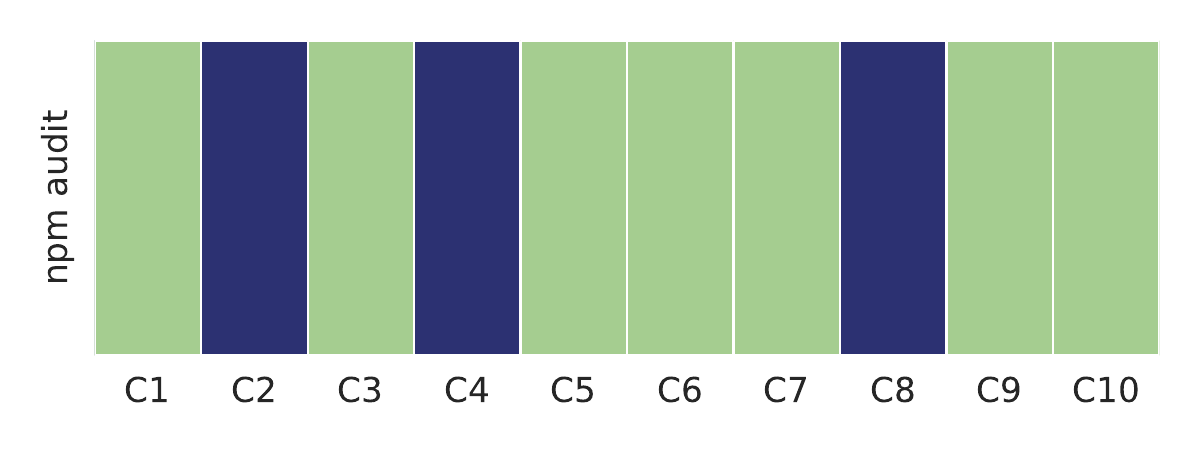}  
  \caption{Control Group: npm audit}
  \label{fig:task1_ctrl}
\end{subfigure}
\caption{Task 1 Vulnerabilities with complexities: Changing of dependency update prioritization after changing the tool from Dependabot to \tool~and \npma}
\label{fig:task1}
\end{figure}

\paragraph{Task 1 Results}
As shown in Figure~\ref{fig:task1_exp}, the number of participants who saw DGV using \tool~and changed their prioritization order is larger than the number of participants who used \npma. 
The list of vulnerabilities provided by the \npma~affected the developers' decision in 3 out of 10 cases, while DGV affected the developers' decision to update vulnerabilities in 7 out of 10 cases~\footnote{The full details of the ranking in Task 1 and Task 2 can be found on our study website:~\url{https://github.com/MUICT-SERU/V-Achilles}}. 
% For the experimental group, the re-prioritize group includes 7 participants (E1, E2, E4, E5, E7, E8, and E9). We observed that they only emphasized severity when they saw the vulnerability report from Dependabot. However, after they used our approach, they also took other factors into account. 
After using \tool, there were six cases (E1, E2, E4, E5, E7, and E8) where the dependency complexity had become one of the factors when prioritizing the package update, either as the first or the second priority factor. 

% \paragraph{\textbf{Qualitative Feedback:}}
In terms of the participant feedback, most observations were related to how the visualization helped them to identify the complexity of the dependencies.
For example, Participant E1 explained as follows: 

\begin{quote}
    \textit{``I added more emphasis on high severity and complex dependency because of its complexity''}.
\end{quote}

Similarly, participant E2 had the same sentiment:

\begin{quote}
    \textit{``[After seeing \tool's visualization] Fixing those dependencies with a larger number of interdependencies and a higher level of severity first and moving to fewer number of interdependencies''}.
\end{quote}

Other participants explained how the graph also helped them to prioritize which updates were important.

\begin{quote}
    \textit{``[After seeing \tool's visualization, I can see the] number of transitive dependencies in each library. If the number is high, it may interrupt other libraries once updated.''} 
\end{quote}

while one participant stated that severity was the only thing they relied on when seeing the Dependabot report.

\begin{quote}
    \textit{``[By using Dependabot] I choose high severity first followed by time (i.e., recently included dependencies first)''}
\end{quote}

However, the same participant mentioned that after seeing the \tool's visualization, the effort required to update might be related to the complexity of the dependencies.

\begin{quote}
\textit{``[By using~\tool] I choose based on the size of the dependency graph. I work on the small one first because it might take a shorter time to fix).''}
\end{quote}

% The unchanged priority group (E3, E6, and E10) did not change their prioritization order. Nonetheless, they all mentioned that after they see our visualization, they realized the information about the package's complexity easier.

% 	\subsubsection{Control Group (npm audit)}
% For the control group, and 
% as shown in Figure \ref{fig:task1_ctrl}, three participants in the re-prioritize group, including C2, C4, and C8, shifted the prioritization order after using \npma. Even though there was no significant change in the prioritization order, the report from Dependabot and \npma provided vulnerability information at different granularity. It affected the prioritization order since these participants used vulnerability information (CVE) as the prioritization criteria. Seven participants in the unchanged priority group gave the same dependency prioritization between Dependabot and \npma.
% After using both Dependabot and \npma, nine participants still used severity as the first priority factor. No participant mentioned dependency complexity when we asked for the reason for re-prioritization.

%Furthermore, we can see from Table~\ref{table:compare-ach-npm} (Task 1), where vulnerable packages have different complexities, 

\textbf{Summary of Task 1:} Navigating complex dependencies helps developers prioritize their updates. Although, initially, all participants cited severity, seven participants instead opted to re-prioritize based on the inter-dependencies compared to the other tools.

% \subsection{Experiment Procedure}
%The experimental group was exposed to only the dependency graph visualization of \tool~without the vulnerability report. 

% \begin{figure*}
% 	\centering
% 	\begin{subfigure}[b]{\textwidth}
% 		\centering
% 		\includegraphics[width=0.5\textwidth]{Figures/screenshot-dependabot.png}
% 		\caption{Dependabot}
% 		\label{fig:screenshot-dependabot}
% 	\end{subfigure}
% 	\hfill
% 	\begin{subfigure}[b]{\textwidth}
% 		\centering
% 		\includegraphics[width=0.5\textwidth]{Figures/screenshot-npm-audit2.png}
% 		\caption{npm audit}
% 		\label{fig:screenshot-npm-audit}
% 	\end{subfigure}
% 	\hfill
% 	\begin{subfigure}[b]{\textwidth}
% 		\centering
% 		\includegraphics[width=0.5\textwidth]{Figures/screenshot-achilles.png}
% 		\caption{Achilles}
% 		\label{fig:screenshot-achilles}
% 	\end{subfigure}
% 	\caption{The security reports of the three tools used in the user study}
% 	\label{fig:screenshots}
% \end{figure*}
	
% \subsection{Analysis}

%\section{Results}\label{sec:Results}
%\subsection{Answering RQ1.}

\begin{figure}[ht]
\begin{subfigure}{\columnwidth}
  \centering
  % include the first image
  \includegraphics[width=\columnwidth]{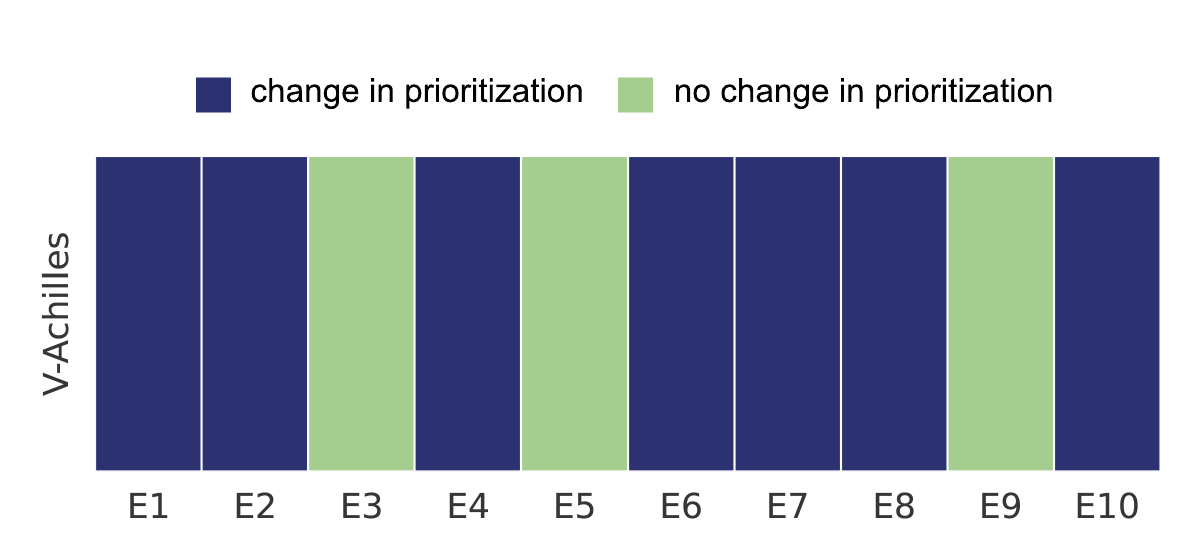}  
  \caption{Experimental Group: V-Achilles}
  \label{fig:task2_exp}
\end{subfigure}
\begin{subfigure}{\columnwidth}
  \centering
  % include the second image
  \includegraphics[width=\columnwidth]{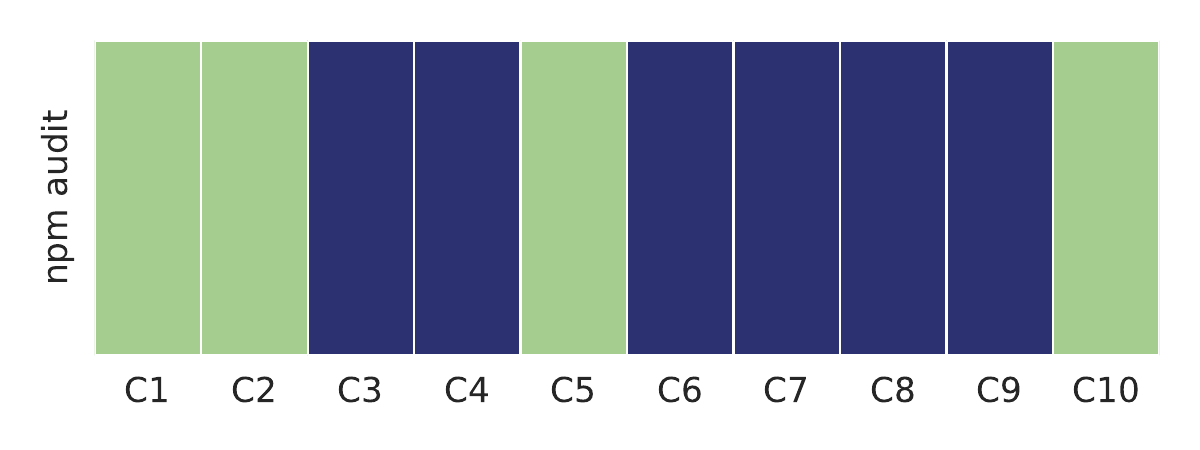}  
  \caption{Control Group: npm audit}
  \label{fig:task2_ctrl}
\end{subfigure}
\caption{Task 2: Changing of dependency update prioritization after changing the tool from Dependabot to V-Achilles and \npma}
\label{fig:task2}
\end{figure}

% \paragraph{Task Two: Navigating Indirect Dependencies}
%\textbf{Task 2: Vulnerabilities with Indirect Dependencies.}~
\paragraph{Task 2 Results}
According to Figure~\ref{fig:task2}, the number of participants who used DGV and changed the prioritization order is also larger than participants who use \npma. 
%The graph visualization in \tool~outperforms npm audit in providing additional information about direct and indirect dependencies although both tools can show report about direct and indirect dependencies. 
However, the differences between the two groups are not as many as in Task 1. 
As shown in Figure~\ref{fig:task2_exp} and Figure~\ref{fig:task2_ctrl}, the list of vulnerabilities provided by \tool's dependency graph visualization affected the developers' decision in 7 out of 10 cases, while the \npma~report affected the developers' decision to update vulnerabilities in 6 out of 10 cases. 
%\raula{please talk about Figure 9b}

Furthermore, we observed that, although the numbers of participants in the re-prioritize group of \tool~and \npma~are close (7 and 6), the participants using \tool~explained that they used the information about direct/transitive dependencies in their reprioritization more than the \npma~group. Some of the feedback from the participants is as follows.

% \raula{please talk about Figure 10}

% \paragraph{\textbf{Qualitative Feedback:}}
Most observations made by all seven participants were related to how the visualization helped them to consider other factors when prioritizing their updates.
For example, when we asked participants to explain their decision before and after seeing the \tool's visualization. Participant E1 said as follows.

\begin{quote}
    \textit{``[After seeing the visualization] I checked their severity and the dependency whether direct or not. \texttt{netmask} and \texttt{base64-url} are high severity but \texttt{netmask} is direct dependency.
I think direct dependency is easier to fix than transitive dependency, then I think it is the highest priority than others.''.} 
\end{quote}

Similarly, another participant (E2), explained that after exposure to the visualization, the participant did consider the transitive dependency as a factor:

\begin{quote}
    \textit{``[After seeing the visualization] I’m prioritizing \texttt{base64-url} and \texttt{minimist} since they are direct vulnerabilities and from the level of severity. For \texttt{netmask} and \texttt{angular-expressions}, thanks to V-Achilles I can see that those libraries are transitive dependencies so I wouldn't be able to actually update those directly and will have to update the direct dependencies instead''.}
\end{quote}

The responses from the participants without using DGV expressed a more diverse range of factors for them to update.
For example, in the control group without exposure (see Figure~\ref{fig:task2_ctrl}), six participants are in the re-prioritize group (C3, C4, C6, C7, C8, and C9) and four participants are in the unchanged priority group (C1, C2, C5, and C10). 
When they saw the vulnerability report from Dependabot, they prioritized severity as the first or second prioritization factor. Participants C3 and C6 used severity as their only factor. 
Participant C4 considered the CVE, and Participant C7 prioritized based on the ease of fixing. Participant C8 only took vulnerability information into account. 
Participant C10 considered the risk of vulnerabilities from attackers. 
However, after they used the \npma, participants C3 and C6 considered the types of vulnerabilities as a factor for prioritization. 
C4, C7, and C8 mentioned that the short description that was provided by the \npma~affected their decision on the prioritization. 
Participant C10 only considered the severity. 

\textbf{Summary of Task 2:} DGV by \tool~allowed users to consider a direct and transitive dependency as a factor when prioritizing when to update. Participants who were exposed to DGV considered the transitivity (indirect dependencies) as a factor compared to the other tools. Seven participants instead opted to re-prioritize their decisions of dependency update.

\textbf{Answer to RQ: Compared to Dependabot and \npma, \tool's DGV influences
the developer’s decision to update the dependencies by taking into account additional information provided by DGV~including (1) the complexity of the dependencies and (2) the transitivity of the vulnerable dependencies.}

\section{Related Work}\label{sec:RelatedWork}

Prior studies show that developers are slow in updating their dependencies \cite{Robbes2012,Hora2015,Sawant2016,Bavota2015,Ihara2017} due to various factors such as compatibility issues, being unaware, and the migration effort outweighing the benefits.
% Another work \cite{Kula2018} similarly finds that most developers were unaware of dependency updates, and also cited other issues such as the migration effort as a barrier to adopting a dependency update.
% Updating the dependencies is an essential task for software project.
% The new release of dependencies introduces new features and fixes to address bugs or security vulnerabilities.
% However, prior studies showed that developers are slow in updating their dependencies.
%The list of related studies is shown below.
\cite{Robbes2012} and \cite{Sawant2016} studied how developers react to deprecated APIs.
They found that only a minority of software projects upgrade their APIs and tend to use the older versions instead.
\cite{Ihara2017} showed that software projects do not always use the latest version of their dependencies.
Decan et al.~\cite{Decan2017} investigated the evolution and issues of dependencies from different ecosystems.
They revealed that the studied ecosystems have an issue with dependency updates, which not only affect the direct users of dependencies but also the indirect users of such dependencies.
\cite{Bogart:2015} conducted a survey and found that developers prefer to keep outdated dependencies as they are afraid the new version might break their code.

% \textbf{Security Vulnerabilities and Updates.}
Security vulnerabilities from third-party dependencies are one of the issues that were raised to the attention of the open-source software communities in recent years~\cite{Web:octoverse}.
There are several studies that investigate the impact of vulnerabilities within different ecosystems (e.g., npm, PyPI)~\cite{Kikas2017,Decan2018,Linares2017,Lauinger2016ThouSN}.
Their results share a similar view that a large number of dependencies were vulnerable regardless of which ecosystem they belonged to.
The important updates, such as vulnerability fixes or new major versions that introduce new features, also influence how long developers would adopt these updates.
\cite{Decan2018} found that the vulnerabilities are prevalent in the dependencies, while the developers of the vulnerable dependencies took several months to release the fixes.
\cite{Chinthanet2021} focused on the lags of updates of vulnerability fixes in the npm ecosystem.
They confirmed that despite the size of vulnerability fixes being small in terms of commits, developers still slowly react to those fixes and cause lags of updates throughout the dependency network.
They also found that the severity of vulnerabilities is one of the factors that influence such lags in updates.
There are related studies to our paper in the literature, which we discuss and compare to our study as follows.
Li et al.~\cite{Li20221} propose an approach called PDGraph. They studied publicly known security vulnerabilities of reused libraries in Java projects and disclosed four underlying risks that can lead to more vulnerabilities in the project. 
The risks include the number of dependency projects, the number of dependent projects, the length of the dependency path, and circular dependencies. 
They studied Java projects, while our study focuses on npm projects, and both of them may have different ecosystems.
DependencyVis~\cite{Lui2021} is a visualization approach for software dependencies that facilitates developers in analyzing the suitability of dependencies in their project by providing various information on basic metrics in the graph visualization including popularity, activeness, vulnerabilities, and license. The study proposes the technique with a prototype tool. However, it does not perform an evaluation. We evaluated our approach using a user study.

Arora et al.~\cite{Arora2016} propose to
use dependency graphs to support collaboration over GitHub. The dependency graphs capture direct and indirect conflicts of concurrent editing of related artifacts in Java Project repositories. 
%The relationships between artifacts in the project are stored in the Neo4j graph database. 
The scope and purpose of this research are different from \tool~which is to analyze the vulnerability dependencies in the npm project.
Wu et al.~\cite{Wu2010} studied the visualizations of the relationship between Windows modules and other binaries. Their approach relied on run-time traces. Users are able to investigate the various aspects of software dependencies between modules and other binaries including dynamically linked libraries on Windows in different aspects. Compared to our work, their approach similarly shows dependencies, but without security vulnerabilities.
Saba Alimadadi~\cite{SabaAlimadadi2013} presents a dependency graph visualization tool called CZSaw. However, the tool does not target software projects, while \tool~is directly applied to npm projects.

\section{Threats to Validity}\label{sec:ThreatsToValidity}
%We now discuss the threats. 
This study has the following threats to its validity.

\textit{Internal Validity:}~A key threat to internal validity is the bias of participants' experiences. Some of the participants may not be familiar with Dependabot or \npma~or may not have performed dependency updates before. We mitigate this threat by providing two example tasks for them to train and understand how the tools work before performing the actual study. 
Moreover, the participants are both students and developers and may have different programming experiences and security awareness. 

\textit{External Validity:}~A key threat to external validity is the generalization of the results in the real world. We simulate the actual dependency prioritization activity by creating the two tasks in the user study. Hence, prioritization results and the factors for prioritization may be limited to the scenarios and vulnerabilities captured by the two tasks. Moreover, the findings are from npm projects and may not be generalized to other ecosystems.

\section{Conclusion}\label{conclusion}
% In this paper, we created a web-based interactive tool called \tool~that analyzes GitHub projects and creates a dependency graph to show direct and transitive dependencies and their vulnerabilities. 
% \tool~allows the developers to explore dependencies in their software intuitively and provides information on vulnerable dependencies to help the developers make decisions on updating such dependencies. 
% % We present the usage scenarios of \tool~and an evaluation of the tool performance with 10 projects on GitHub. 
% The future work includes performing an evaluation of the dependency graph visualization with developers to assess its effectiveness in assisting their decision of dependency updates. We plan to also improve \tool~to detect vulnerabilities in a longer chain of transitive dependencies (the tool currently supports 4 levels of transitive dependencies). An improved visualization method may be needed due to the complexity of the graph after analyzing all the transitive dependencies.
% Moreover, visualization like \tool~can also be included in the code review process to assist the reviewers with the security of the added packages.

This paper studies the effectiveness of a dependency graph visualization 
to motivate developers to update vulnerable dependencies.
Using \tool~tool that analyzes GitHub projects and creates a dependency graph visualization to show direct and indirect dependencies and their vulnerabilities, we performed a user study with 20 participants to evaluate our approach. 

The results show that 7 out of the 10 participants who used our visualization after Dependabot changed their prioritization in the two tasks of a project with vulnerable complex dependencies and a project with vulnerable direct and indirect dependencies. Only 3 out of 10 participants and 6 out of 10 participants used \npma~after Dependabot changed their prioritization in both tasks respectively. 
%We also found that the participants who used the \npma~also changed their prioritization after seeing our visualization. 

This study encourages the incorporation of dependency graph visualization into modern dependency management tools to aid developers in making dependency update decisions and more research on adopting new ways of visualizations for dependency management.
The data of the empirical evaluation is publicly available at~\url{https://github.com/MUICT-SERU/V-Achilles}.

\section*{Acknowledgement}
This work was financially supported by the Office of the Permanent Secretary, Ministry of
Higher Education, Science, Research and Innovation (OPS MHESI) Grant No. RGNS 64-143, and the Faculty of Information and Communication Technology, Mahidol University, Thailand.

\bibliographystyle{IEEEtran}
\bibliography{references}

\end{document}